

\magnification=1200


\def\Rrm{\hbox{\rm I\hskip -2pt R}}
\def\Nrm{\hbox{\rm I\hskip -2pt N}}
\def\Crm{\hbox{\rm l\hskip -5.5pt C\/}}


\font\sevenit=cmti7 \relax


\def\psaut{\vskip 5pt plus 1pt minus 1pt}
\def\saut{\vskip 10pt plus 2pt minus 3pt}
\def\gsaut{\vskip 20pt plus 3pt minus 4pt}

\def\sqr#1#2{{\vcenter{\vbox{\hrule height.#2pt
\hbox{\vrule width.#2pt height#1pt \kern #1pt
\vrule width.#2pt}
\hrule height.#2pt}}}}

\def\fl{\rightarrow}


\centerline {\bf Topological Quantum Groups, Star Products}
\psaut
\centerline {\bf and their relations}
\saut
\centerline {Mosh\'e Flato and Daniel Sternheimer}

\centerline {Physique Math\'ematique, Universit\'e de Bourgogne}

\centerline {F-21004 DIJON Cedex, FRANCE}

\centerline {\sevenrm (flato@satie.u-bourgogne.fr, daste@ccr.jussieu.fr)}

\saut
\centerline{\it Dedicated to our friend Ludwig Faddeev on his 60th birthday.}
\gsaut
\centerline {\bf Abstract.}
\sevenrm
{This short summary of recent developments in quantum compact groups and star
products is divided into 2 parts. In the first one we recast star products
in a more abstract form as deformations and review its recent developments.
The second part starts with a rapid presentation of standard quantum group
theory and its problems, then moves to their completion by introduction of
suitable Montel topologies well adapted to duality. Preferred deformations
(by star products and unchanged coproducts) of Hopf algebras of functions
on compact groups and their duals, are of special interest. Connection with
the usual models of quantum groups and the quantum double is then presented.}
\tenrm
\gsaut
\centerline   {\bf 0 - INTRODUCTION.}
\psaut
The idea that quantum theories are deformations of classical theories was
presumably in the back of the mind of many scientists, even before the
mathematical notion of deformation was formalized by Gerstenhaber [G]
for algebraic structures. We were even told by witnesses (many of whom
contribute to this volume) that Ludwig Faddeev mentioned that idea in his
lectures on quantum mechanics in Leningrad in the early 70's, around the
time when the so-called geometric quantization was developed.

However in all these approaches people were always considering that in the end
quantum theories have to be formulated in operator language, while an essential
point in our approach ([FS1], [B{\it ea}]) is that quantum theories can be
developed in an autonomous manner on the algebras of classical observables by
deforming the algebraic structures. The connection with operatorial
formulation, whenever possible, comes only afterwards and is optional.
This applies both to quantum mechanics and quantum field theories.
Our approach is often referred to as star-products, or deformation
quantization.

Around the beginning of the 80's, when it became rather clear that constructive
quantum field theory (at least in 4 dimensions) was facing tremendous
analytical problems, the school of Faddeev tried a new approach to quantization
of field theories, first with 2-dimensional integrable models. Doing so they
discovered [KR] the beginning of what turned to be [FRT] a mathematical gold
mine, to which both mathematicians and theoretical physicists rushed
(and the rush is still in full speed): quantum groups.

In this short Note we shall present both theories in a context that makes the
relations between both quite natural. This presentation (especially its second
part) relies on a paper [BFGP]) now being published, where necessary details
can be found.
\gsaut
\centerline {\bf 1 - DEFORMATIONS, QUANTIZATIONS AND STAR PRODUCTS.}
\psaut
{\bf 1.1. The framework.} Let $A$ be an algebra. In the following it can be an
associative algebra (vector space with product and unit), a Lie algebra, a
bialgebra (associative algebra with coproduct), a Hopf algebra (bialgebra with
counit and antipode), etc., with the usual compatibility relations between
algebraic laws. For simplicity of notations we shall take the base field to be
$\Crm$ (the complex numbers). It can also be a topological algebra, i.e. any
of the above when the vector space is endowed with a topology such that all
algebraic laws are continuous mappings. We shall specify the kind of algebra
considered whenever needed.

An example of such an algebra is given by the Hopf algebra $\hskip.1cm
\Crm [t]$ of complex polynomials in one variable $t,$ with product
$t^n \times t^p = \biggr(\matrix {n+p \cr p \cr }\biggl) \hskip.1cm
t^{n+p},$ coproduct $\delta (t^n) = \displaystyle{\sum^n_{i=0}}
\hskip.1cm t^i \otimes t^{n-i},$ counit $\varepsilon (t^n) = \delta_{n 0}$
(Kronecker $\delta$) and antipode $S(t^n) = (-1)^n \hskip.1cm t^n.$

Its dual (in a sense we shall make precise in the following) is the bialgebra
of formal series $\Crm [[t]],$ with usual product and coproduct given by
$\Delta f (t,t') = f(t+t') \in \Crm [[t,t']]$ for $f \in \Crm [[t]].$

Now if we extend the base field to the ring $\hskip.1cm \Crm [[t]],$ we get
from $A$ the module $\tilde{A} =
A[[t]]$ of formal series in $t$ with coefficients in $A,$ on which
we can consider algebra structures.
\psaut
{\bf 1.2. Definition.} \it A deformation of an algebra $A$ is a (topologically
free in the case of topological algebras) $\hskip.1cm \Crm [[t]]$ algebra
$\tilde{A}$ such that the quotient of $\tilde{A}$ by the ideal
$t \tilde{A}$ generated by $t$ is isomorphic to $A.$ \rm

For an associative algebra this means that on $\tilde{A}$ there is a new
product, denoted by $*$, such that for $a,b \in A$ :
$$a * b = \sum^\infty_{r=0} \hskip.1cm t^r \hskip.1cm C_r (a,b) \eqno{(1)}$$
where $C_0 (a,b) = ab$ (the product of $A$), and the cochains $C_r \in
{\cal L} (A \hat{\otimes} A, A),$ the space of linear (continuous) maps
 from the (completed, for some adequate topology, in the topological case)
 tensor product $A \otimes A$ into $A.$ The associativity condition
 for $*$ gives as usual [G] conditions on the cochains $C_r$ (e.g. $C_1$
 is a cocycle for the Hochschild cohomology).

For a Lie algebra one has similar relations (with Chevalley cohomology),
and for bialgebras an adequate cohomology can be introduced [B1].

For a bialgebra, denoting by $\otimes_t$ the tensor product of
$\hskip.1cm \Crm [[t]]$ modules, one can identify $\tilde{A}
\hat{\otimes}_t \tilde{A}$ with ($A \hat{\otimes} A)
[[t]]$ and therefore the deformed coproduct is defined by
$$\tilde{\Delta} (a) = \sum^\infty_{r=0} \hskip.1cm t^r D_r (a),
\hskip.1cm a \in A \eqno{(2)}$$
where $D_i \in {\cal L} (A, A \hat{\otimes} A)$ and $D_0$ is the
coproduct $\Delta$ of $A.$

For a Hopf algebra, the deformed (Hopf) algebra has same unit and counit,
but in general not the same antipode.

As in the algebraic theory [G], two deformations are said {\it equivalent}
if they are isomorphic as $\Crm [[t]]$ (topological) algebras, the
isomorphism being the identity in degree $0$ (in $t$).
And a deformation $\tilde{A}$ is said {\it trivial} if it is equivalent
to the deformation obtained by base field extensions from the algebra $A.$
\psaut
{\bf 1.3. Example. Star products.} We take $A = C^\infty (W),$ with $W$ a
symplectic (or Poisson) manifold with $2$-form $\omega$. On $A$ we have
a Poisson bracket $(a,b) \mapsto P(a,b),$ which is a bidifferential
operator of order (1,1). We say that (1) defines a {\it star product} on
the associative algebra $A$ (with pointwise multiplication) if in addition:
$$C_1 (a,b) - C_1 (b,a) = 2P (a,b) \hskip.3cm a,b \in A. \eqno{(3)}$$
We do not assume here that the $C_r$ are bidifferential operators,
nor n.c. (null on constant functions, which implies that the function 1 is
a unit for the deformed algebra as well). If we do, then [B{\it ea}]
it is coherent to restrict oneself to the corresponding Hochschild
cohomologies. But in star representations (see below)
one encounters often bipseudodifferential cochains $C_r.$

{}From (3) follows that the star product defines a deformation of the
Lie algebra $(A,P)$ by :
$$[a,b]_* \equiv {1 \over 2t} (a*b-b*a) = P(a,b) + \sum^\infty_{r=2}
\hskip.1cm {1 \over 2} t^{r-1} (C_r (a,b) - C_r (b,a)). \eqno{(4)}$$
This allows (in the differentiable case) to use instead of the
infinite-dimensional Hochschild cohomologies, the finite-dimensional
Chevalley cohomology spaces. E.g. the dimension of Chevalley $2$-cohomology
is (in the n.c. case) $1+b_2 (W)$ where $b_2 (W)$ is the second Betti
number of $W$ which permits (as in [B{\it ea}]) to show that at
each level there are only $1+b_2 (W)$ choices.
\psaut
{\bf 1.4. Typical example : Moyal on $\Rrm^{2n}$.} In 1927, H. Weyl [W] gave
 a rule for passing from a classical observable $a \in A = C^\infty
 (\Rrm^{2l})$ to an operator in $L^2 (\Rrm^l)$ which represents a quantization
 of this observable. It can be written
$$A \ni a \mapsto \Omega_w (a) = \int \tilde{a} (\xi, \eta) \hskip.1cm
exp (i(P \xi + Q \eta) / \hbar) \hskip.1cm w (\xi, \eta) \hskip.1cm
d^l \xi \hskip.1cm d^l \eta \eqno{(5)}$$
where $\tilde{a}$ is the inverse Fourier transform of $a,$ $P$ and $Q$
satisfy the canonical commutation relations $[P_\alpha, Q_\beta] = i \hbar
\hskip.1cm \delta_{\alpha \beta} \hskip.1cm (\alpha, \beta = 1,...,l),$ $w$
is a weight function ($=1$ in the case of Weyl) and the integral is taken
in the weak operator topology. An inverse formula was given a few years
later by E. Wigner [Wi], and numerous variants exist.
Whenever either side is defined, the trace can be given by:
$$Tr (\Omega_1 (a)) = (2 \pi \hbar)^{-l} \hskip.1cm \int_{\Rrm^{2l}}
\hskip.1cm a \hskip.1cm \omega^l \eqno{(6)}$$
In the end of the 40's, starting from a point of view different from ours,
Moyal [M] and Groenewold [Gr] found that the commutator and product (resp.)
of quantum observables correspond, in the Weyl rule, to sine and exponential
of the Poisson bracket (resp.), with the parameter $t = {1 \over 2}
\hskip.1cm i \hbar.$ Thus $\Omega_1 (a) \hskip.1cm \Omega_1 (b) =
\Omega_1 (a *_M b)$ where $*_M$ is given by (1) with (for $r \geq 1$)
$r ! C_r (a,b) = P^r (a,b),$ the $r^{th}$ power of the bidifferential
operator $P.$
\psaut
{\bf 1.5. Quantizations.}
 In 1975, inspired by our earlier works [FLS] on $1$-differentiable
deformations of the Lie algebras $(A,P),$ J. Vey [V] obtained what turned
to be the Moyal bracket as an example of differentiable deformation,
and showed its existence on any symplectic $W$ with $b_3 (W) = 0.$
We then not only made the connection with quantization but also showed,
with examples, that quantization should in fact be considered as a
deformation of a classical theory, with the same algebra of observables
and a star-product [B{\it ea}]. Around the same time and independently,
Berezin [B] had shown that the normal ordering of physicists (weight
$w(\xi, \eta) = exp (- {1 \over 4} (\xi^2 + \eta^2))$ in (5)) can be
defined for more general manifolds than $\Rrm^{2l}.$ That ordering is
the analogue (for complex coordinates $\xi \pm i \eta$) of the standard
ordering (weight $w(\xi, \eta) = exp ( - {1 \over 2} i \xi \eta))$
which mathematicians are using in pseudodifferential
operator theory, and is preferred for field theory quantization.

In our approach, we have an autonomous definition of the spectrum of an
observable. To that effect we consider the star exponential (the analogue
of the evolution operator)
$$Exp (sa) = \sum^\infty_{n=0} \hskip.1cm {1 \over n!} \hskip.1cm s^n
(i \hbar)^{-n} \hskip.1cm (a*)^n \eqno{(7)}$$
(the sums involved being taken in the distribution sense) and define
the spectrum of the observable $a$ to be that (in the sense of L. Schwartz)
of the star exponential distribution, i.e. the support of its Fourier-Stieltjes
transform (in $s$). For the harmonic oscillator for instance, one gets
$(n + {1 \over 2} l) \hbar$ with Moyal ordering $(n \in \Nrm)$ and
$n \hbar$ with normal ordering (which explains why it is favored when
$l \fl \infty$). But many other examples can be treated,
e.g. the hydrogen atom with $W = T^\forall \hskip.1cm S^3$ for manifold.

Star products can also be defined when dim $W = \infty,$ and there one
can e.g. find some cohomological cancellations of infinities [Di] by taking
orderings "in the neighbourhood of normal ordering": this amounts to
substracting an infinite coboundary from an infinite cocycle to get
a finite ("renormalized") cocycle.
\psaut
{\bf 1.6. Closed star products.} Whenever there is a (generalized) Weyl
mapping between $A = C^\infty (W)$ (plus possibly some distributions,
or part of it only) and operators on a Hilbert space (typically a space of
square integrable functions in "half" of the variables, via some polarization),
some of these operators will have a trace. Therefore it is natural to ask
whether a functional with the properties of a trace can be defined
on the algebra $(A, *).$

For Moyal ordering one has (6). For other orderings on $\Rrm^{2l}$ that
formula is valid modulo higher powers of $\hbar.$ Therefore [CFS]
a natural requirement is to look at the coefficient of $\hbar^l$
in $a*b,$ where $a,b \in A[[\hbar]],$ and require that its integral
over $W$ is the same as that of $b*a.$ Or equivalently :
$$\int_W \hskip.1cm C_r (a,b) \hskip.1cm \omega^l = \int_W
\hskip.1cm C_r (b,a) \hskip.1cm \omega^l \eqno{(8)}$$
whenever defined for $a,b \in A$ and $1 \leq r \leq l.$ A star-product (1)
satisfying (8) is called {\it closed}. If (8) is true for all $r$ we call it
{\it strongly closed}. Note that, in view of (3), (8) is always true
 for $r=1$ -- so that all star products on $2$-dimensional manifolds
 are closed. It has been shown by Boris Tsygan (as a consequence of the
 definition of the trace, in [NT]) that all differentiable n.c. star products
 are equivalent to strongly closed ones. (There exist however non closed
 star products, that e.g. are not null on constants).

An interesting feature of closed star products [CFS] is that they are
classified by {\it cyclic} cohomology [C], instead of only Hochschild
cohomology. This suggests to define, in parallel to the similar notion
for operator algebras [C], the {\it character} of a closed star-product
as a cocycle $\varphi$ in the cyclic cohomology bicomplex with components
(non zero only for $l \leq 2k \leq 2l$) :
$$\varphi_{2k} (a_0, a_1,..., a_{2k}) = \int_W \hskip.1cm a_0 * \tau (a_1,a_2)
*...* \tau (a_{2k-1}, a_{2k}) \hskip.1cm \omega^l \eqno{(9)}$$
where $\tau (a,b) = a*b - ab$ measures the noncommutativity of the star
product.
It can be shown [CFS] that for $W = T^* M, M$ compact Riemannian manifold, and
for the star product of standard ordering (composition of symbols of
pseudodifferential operators), the character coincides with that given by
the trace on pseudodifferential operators. Therefore, using the algebraic
index theorem of [CM], it is given by the Todd class Td$(T^*M)$ as a
current over $T^*M.$
\psaut
{\bf 1.7. Existence.} Jacques Vey [V] had obtained the existence of
 star brackets for all symplectic manifolds with $b_3 = 0,$ and this
 was extended ([NV], [L]) to star products (under the same hypothesis).
 The underlying idea is to "glue" Moyal products on Darboux charts, and the
 condition $b_3 = 0$ is needed to control multiple intersections of charts.
 But we knew from the beginning [B{\it ea}] that this condition is not
 necessary. Then M. Cahen and S. Gutt showed existence for $W = T^*M,$
$M$ parallelisable, and soon afterwards [LDW1] existence was shown for any $W$
symplectic (or regular Poisson) manifold.

In 1985-86 (in obscure form, made more clear only recently) B. Fedosov [F]
gave a geometrical and algorithmic construction of star products on any
$W$ by viewing $A[[t]]$ as a space of flat sections in the bundle of
(formal) Weyl algebras on $W$ (and pulling back the multiplication of
sections; a flat connection on that bundle is algorithmically constructed
starting with any symplectic connection on $W$). The geometric background
of Fedosov's construction has been recently explicited further by several
authors ([Gu], [EW]).

Using also Weyl algebras, but here essentially [LDW2] to build
compatible local equivalences that allow to "glue together" Moyal products
a Darboux charts, it has been possible [OMY] to give another and more concrete
proof of existence of star products on any $W,$ and even to do it in a way
that proves directly also existence of closed star products.
\psaut
{\bf 1.8. Star representations.} When $a$ is a generator of a Lie algebra
 ${\cal G}$ of functions (e.g. on a coadjoint orbit of a Lie group $G$),
 the star exponential (7) gives the corresponding one-parameter group.
 And if the star commutator (4) coincides, for $a,b \in {\cal G},$
 with $P(a,b),$ the Poisson bracket (which is the Lie bracket in this case),
 one can (by the Campbell-Hausdorff-Dynkin formula) generate a realization
 of $\tilde{G}$ (the connected and simply connected Lie group with Lie algebra
 ${\cal G}$) by the star exponentials (7) and their star products.
Such a star product is said {\it covariant}.

It is said {\it invariant} if $[a,b] = P(a,b) \hskip.1cm \forall a
\in {\cal G}$ and $b \in A$ (this is the geometric invariance of the
star product under the action of $G$). There do not always exist invariant
star products (e.g. for nilpotent groups of length $>2$), but covariant
ones always exist. For covariant star products, the geometric action of
$G$ is modified by a $t$-dependent multiplier.

We call {\it star representation} the distribution on $G$ defined by the
star exponential associated with a covariant star product. Such
representations have been built for all compact and all solvable Lie groups,
some series of representations of semi-simple groups (including some of
those with unipotent orbits), and other examples. The cochains $C_r$
obtained here are in general pseudodifferential.
\gsaut
\centerline {\bf 2 - TOPOLOGICAL QUANTUM GROUPS.}
\psaut
{\bf 2.1. The setting.} Let $G$ be a Poisson-Lie group, i.e. a Lie group with
Poisson structure, such that for the usual coproduct $\Delta$ on the Hopf
algebra $H = C^\infty (G),$ (i.e. $\Delta a(g,g') = a(gg') \hskip.1cm ;
\hskip.1cm g,g' \in G$), the Poisson bracket $P$ (on $G$ or $G \times G$)
satisfies
$$\Delta \hskip.1cm P(a,b) = P(\Delta a, \hskip.1cm \Delta b)
\hskip.3cm a,b \in  H \eqno{(10)}$$

Equivalently we can consider the Lie bialgebra ${\cal G}$;
the dual ${\cal G}^*$
has a bracket $\varphi^* : {\cal G}^* \wedge {\cal G}^* \fl {\cal G}^*$ such
that its dual $\varphi$ is a $1$-cocycle for the adjoint action.
When $\varphi$ is the coboundary of some $r \in {\cal G} \wedge
{\cal G}$ (solution of the classical Yang-Baxter equation) it is said that
 the Poisson-Lie group is triangular. In that case there exists a $G$-invariant
 differentiable star product on $H,$ and the associativity condition for that
 star product gives a solution to the quantum Yang-Baxter equation:
 the deformed algebra $H$ is the realization of a quantum groups [D].
 Furthermore [T] there exists a (non-invariant) equivalent star product
$*'$ on $H$ such that (for the same $\Delta$ as above)
$$\Delta (a*'b) = \Delta a *' \Delta b \eqno{(11)}$$
and the same for the commutator, which is clearly a quantization of (10).

In the "dual" approach of Jimbo [J], one deforms $\Delta$ to some $\Delta_t$
on some completion ${\cal U}_t ({\cal G})$ of the enveloping algebra
${\cal U} ({\cal G}).$  It is this deformation  that was first discovered [KR],
for ${\cal G} = sl(2)$: the commutation relations which define ${\cal U}_t$
have a deformed form (one of them becoming a sine instead of a linear
function).

In line with our philosophy, it is thus natural to ask whether the deformed
algebra ${\cal U}_t$ can be realized (instead of an operatorial realization)
by classical functions and some star product giving the deformed commutators.
It turns out that this is possible [FS], with a star-product using a new
parameter $\hbar$ unrelated to $t.$ In fact, since there is some duality
between $H$ and ${\cal U}_t$ (we shall make this more precise later),
the two parameters $t$ and $\hbar$ are in a way dual one to the other:
the deformed algebra $H[[t]]$ (with star product) gives a deformed coproduct
on ${\cal U}_t$ that induces deformed commutation relations expressible
with another star product (with a new parameter $\hbar$).
Moreover the latter expression is essentially unique [FS2] due to a strong
invariance property that essentially characterizes the star-sine for the
Moyal star product. These star realizations (with $\hbar$) can be given
([Lu], [FLuS]) for various series of classical Lie algebras.

We have just seen that duality plays an important r\^ole in the Hopf
algebraic formulation of quantum groups. But there is a fundamental
difficulty, that until recently was quietly avoided : the algebraic dual
of an infinite-dimensional Hopf algebra $A$ is not Hopf and the bidual
is strictly larger than $A.$ So (unless $G$ is a finite group !) one has
to be extremely careful in dualizing - or topologize in a suitable fashion.
\psaut
{\bf 2.2. Topological quantum groups : the classical case [BFGP].}
\psaut
{\bf a. Definition.} A topological algebra (resp. bialgebra, Hopf algebra)
$A$ is said {\it well behaved} if the underlying (complete) topological vector
space is nuclear and either Fr\'echet (F) or dual of Fr\'echet (DF) [Tr].

The topological dual $A^*$ is then also well-behaved, and the bidual
$A^{**} = A.$ This is the case when $A$ has countable dimension, with the
strict inductive limit of finite-dimensional subspaces as topology.
For example, $A = \Crm [t]$ (the polynomials) is well-behaved,
and so is $A^* = \Crm [[t]].$

{\bf b. The models.} Let $G$ be a compact connected Lie group.
 Then $H(G) = C^\infty (G)$ and its dual $A(G) = {\cal D}' (G)$
 (the distributions) are well-behaved topological  Hopf algebras.

Now $G$ can be imbedded in ${\cal D}' (G)$ as Dirac distributions at points of
$G,$ and its linear span is dense in ${\cal D}' (G).$ The product on
${\cal D}' (G)$ is the convolution of (compactly supported) distributions,
and the coproduct is defined by $\Delta (x) = x \otimes x$ for $x \in G$
(considered as a Dirac distribution).

We know that the enveloping algebra ${\cal U} ({\cal G})$
can be identified with differential operators on $G,$ i.e. all distributions
with support at the identity. Its "completion" ${\cal U}_t$ will involve
some entire functions of Lie algebra generators, i.e. an infinite sum of
Dirac $\delta$'s and their derivatives, and thus take us outside
${\cal D}'.$ In order to include this model as well one will therefore
have to restrict oneself to a subalgebra of $H.$ The natural choice is the
space ${\cal H} (G)$ of $G$-finite vectors of the regular representation,
which is generated by the coefficients (matrix elements) of the irreducible
(unitary) representations. Thus
${\cal H} (G) = \displaystyle{\sum_{\rho \in \hat{G}}}
\hskip.1cm {\cal L} (V_\rho),$ where $V_\rho$ is the space on which the
representation $\rho \in \hat{G}$ is realized. Its dual is then
$${\cal H}^* (G) = {\cal A} (G) = \prod_{\rho \in \hat{G}} \hskip.1cm
{\cal L} (V_\rho) \supset {\cal D}' (G). \eqno{(11)}$$
The imbedding ${\cal U} ({\cal G}) \ni u \mapsto i(u) = (\rho (u))
\in {\cal A} (G)$ has a dense image for the topology of ${\cal A}$
(the image is of course in ${\cal D}' (G),$ but is {\it not} dense
for the ${\cal D}'$ topology).
\psaut
{\bf 2.3. Topological quantum groups : the deformations.}
\psaut
We shall restrict ourselves here to a summary of the main notions and results
of the theory in the framework explained before, referring to [BFGP] and
references quoted therein for more details.

Duality and deformations work very well together in our setting.
More precisely:
\psaut
{\bf Proposition 1.} \it Let $\tilde{A}$ be a bialgebra (resp. Hopf)
 deformation of a well-behaved topological bialgebra (resp. Hopf algebra)
 $A.$  Then the $\hskip.1cm \Crm [[t]]$ dual $\tilde{A}^*_t$ is a
 deformation of the topological Hopf algebra $A^*.$  Two deformations
 $\tilde{A}$ and $\tilde{A}'$ of $A$ are equivalent iff $\tilde{A}^*_t$
and $\tilde{A}'^*_t$ are equivalent deformations of $A^*.$ \rm

The known models of quantum groups lead us to select a special type
of deformations:

{\bf Definition} (see also [GS]). \it A deformation of the bialgebra
 ${\cal H} (G)$ (resp. $C^\infty (G)$) with unchanged coproduct is called
 a preferred deformation. \rm

This definition is motivated by the following :

{\bf Proposition 2.} \it Let $({\cal H} [[t]], * , \tilde{\delta})$ be a
coassociative deformation of the bialgebra ${\cal H}.$ Then,
up to equivalence, one can assume that $\tilde{\delta} = \delta$
(the coproduct in ${\cal H}$); the product is quasi-commutative and
quasi-associative, the counit unchanged, and if the product is associative
then ${\cal H} [[t]]$ is a $\Crm [[t]]$ Hopf algebra with same unit and
counit as ${\cal H}.$ The same holds for $H.$ \rm

(By quasi-associativity, etc., we means as usual that the associativity, etc.,
condition is satisfied up to a factor). That result is proved by using duality
from the following results for the duals ${\cal A} (G) = {\cal H} (G)^*$ and
$A (G) = H (G)^* = {\cal D}' (G)$ :

{\bf Theorem 1.} \it Let $A$ be either ${\cal A} (G)$ or $A(G).$ Then any
associative algebra deformation of $A$ is trivial, and $A$ is rigid
in the category of bialgebras; any associative bialgebra
deformation of $A$ is quasi-cocommutative and quasi-coassociative. \rm

   More specifically $H^n (A,A) = 0 \hskip.2cm \forall n \geq 1 $ and
$ H^1 (A, A \hat{\otimes} A) = 0 $ (for algebraic and continuous Hochschild
cohomologies), which shows the rigidity of $A$ as bialgebra in the
sense of [B1].
Moreover, if $(A [[t]], \tilde{\Delta})$ is an associative bialgebra
deformation of $A$ with unchanged product, then there exists $\tilde{P} \in
(A \hat{\otimes} A) [[t]]$ such that $\tilde{\Delta} = \tilde{P} \hskip.1cm
\Delta_0 \hskip.1cm \tilde{P}^{-1}$ (where $\Delta_0$ is the coproduct in $A$),
the counit is unchanged, and there exists an antipode $\tilde{S}$ for
$A[[t]]$ that is given by $\tilde{S} = \tilde{a} \hskip.1cm S_0
\hskip.1cm \tilde{a}^{-1}$ where $S_0$ is the antipode of $A$ and
$\tilde{a}$ is some element in $A[[t]].$
Our topological notion of duality also gives us, automatically, that
the deformed product $*$ on the topological dual $H$ (either ${\cal H} (G)$
or $C^\infty (G)$) of $A$ is a {\it star product} (starting with the Poisson
bracket) in the sense of part 1, for all $G$ compact.

In addition, the restriction of a Hopf deformation of $H(G)$ defines a Hopf
deformation of ${\cal H} (G).$ If $\Gamma$ is a normal subgroup of $G,$ any
preferred deformation of ${\cal H} (G)$ gives a preferred deformation of
${\cal H} (G/\Gamma)$ (and the same with $H(G)$): we can define
{\it quotient deformations}, a useful notion e.g. to pass from
 $SU(2)$ to $SO(3)$, etc.
\psaut
{\bf 2.4. Topological quantum groups : the models.}
\psaut
We shall now explain how the known models of quantum groups relate to the
general framework presented in the previous section.

{\bf a. Generators of ${\cal H} (G).$} The algebra ${\cal H} (G),$ $G$ compact,
is a finitely generated domain. We say that a set $\{ \pi_1,...,\pi_r \}
\subset \hat{G}$ of irreducible representations (irrep.) is {\it complete}
if its coefficients
generate ${\cal H} (G).$ For $SU(n), SO(n)$ and $Sp(n),$ the standard
representation is in itself a complete set. For $Spin (n),$ we take the
irreducible spin representation(s) (one for $n$ odd, $2$ for $n$ even).
For $E_6$ (resp. $E_7$) there exist(s) two (resp. 1) irrep. that form
a complete set. For all other exceptional (simply connected compact)
groups, any irrep. is a  complete set.

Define $\pi_0 = \displaystyle{\oplus^r_{i=1}} \hskip.1cm \pi_i,$
and call $\{ C_{ij} \}$ the coefficients of $\pi_0$ in a given fixed basis:
they form a topological generator system for the preferred Hopf deformation
$({\cal H} [[t]], * )$ of ${\cal H}.$ The quasi-commutativity of that
deformation can then be expressed as follows: if $T$ is the matrix $[C_{ij}],
T_1 = T \otimes Id, \hskip.1cm T_2 = Id \otimes T,$ there exists an invertible
$R$ in ${\cal L}$ ($V_{\pi_0} \otimes V_{\pi_0}) [[t]]$ such that
$R \hskip.1cm (T_1 \hskip.1cm * T_2) =
 (T_1 \hskip.1cm * \hskip.1cm T_2) \hskip.1cm R.$

{\bf b. The Drinfeld models} [D1]. Let ${\cal U} = {\cal U} ({\cal G})$ be
the enveloping algebra. Drinfeld has shown [D2] that it is rigid (as algebra),
and there exists a Hopf deformation ${\cal U}_t$ of ${\cal U}$ (endowed with
its natural topology) that is a topologically free complete
$\Crm [[t]]$-module: there is an isomorphism $\tilde{\varphi}:
{\cal U}_t \simeq {\cal U} [[t]]$ as $\Crm [[t]]$-modules, and also as
 algebras; we call such a $\tilde{\varphi}$ a {\it Drinfeld isomorphism}.
The coproduct $\tilde{\Delta}$ of ${\cal U}_t$ is obtained from the original
coproduct by a twist : $\tilde{\Delta} = \tilde{P} \hskip.1cm \Delta_0
\hskip.1cm P^{-1}$ for some $\tilde{P} \in {\cal U}_t \hat{\otimes}_t
\hskip.1cm {\cal U}_t.$

Using the fact that ${\cal U} ({\cal G}) \subset A (G) \subset {\cal A} (G)$
we can extend the Hopf deformation ${\cal U}_t$ to a Hopf deformation of $A(G)$
or ${\cal A} (G)$ with unchanged product, unit and counit. By $\Crm [[t]]$
duality this gives a preferred deformation of $H(G)$ or ${\cal H} (G)$ (resp.).

All this construction depends on the choice of a Drinfeld isomorphism
$\tilde{\varphi},$ but in an inessential way: two Drinfeld isomorphisms
$\tilde{\varphi}$ and $\tilde{\psi}$ give equivalent preferred deformations
of ${\cal H} (G).$ Note that the above $ R$-matrix can be specified to be a
solution of the Yang-Baxter equation.

{\bf c. The Faddeev-Reshetikhin-Takhtajan models.} These [FRT] models are
 recovered by a good choice of the Drinfeld isomorphism: if $\tilde{\rho}$
 is a representation of ${\cal U}_t$ and $\pi = \rho_0 \in \hat{G}$ is its
 classical limit, then there is a Drinfeld isomorphism $\tilde{\varphi}$
 such that $\tilde{\rho} = \pi \circ \tilde{\varphi}.$

When we apply this to $G = SU(n), SO(n)$ or $Sp(n)$ we recover the [FRT]
quantizations of these groups as preferred Hopf deformations of ${\cal H}(G)$
that extend to preferred Hopf deformations of $C^\infty (G).$

{\bf d. The Jimbo models} [J]. These models are somewhat special, because
 we get here nontrivial deformations. We shall explain it here for the case
 ${\cal G} = sl(2).$ The general case is similar, the main difference being
 that there ${\cal U} ({\cal G})$ is extended by Rank$({\cal G})$ parities.

Consider the quantum algebra $A_t$ generated by 4 generators $\{ F,F',S,C \}$
with relations :
$$[F,F'] = 2 SC, \hskip.1cm FS = (S \hskip.1cm {\rm cost} - C) \hskip.1cm F,
\hskip.1cm FC = (C \hskip.1cm {\rm cost} + S \hskip.1cm {\rm sin}^2 t)
\hskip.1cm F \eqno{(12a)}$$
$$F'S = (S {\rm cost} + C) \hskip.1cm F', \hskip.1cm F'C = ({\rm cost} - S
\hskip.1cm {\rm sin}^2 t) F', \hskip.1cm C^2 + S^2 {\rm sin}^2 t = 1,
\hskip.1cm [S,C] = 0. \eqno{(12b)}$$
The more familiar form is obtained by setting $q = e^{it}$ $(t \notin 2
\pi {\bf Q})$ and $S = {K-K^{-1} \over q-q^{-1}}, \hskip.1cm C =
{1 \over 2} (K+K^{-1})$ for some new generators $K$ and $K^{-1}.$
 But we prefer (12) because it is not singular at $t=0,$ and we can thus
 define $\tilde{A}_t$ as the $\Crm [[t]]$ algebra $A_t$
when $t$ is a formal parameter. The usual commutation rules of $sl(2)$
are obtained with $SC, FC$ and $F'C$; therefore $A_0 \simeq {\cal U}
(sl(2)) \otimes P$ where $P \simeq \Crm [x] / (x^2 - 1)$ is generated
by a parity $C$ $(C^2=1$ when $t=0).$

The formal algebra $\tilde{A}_t$ is thus a deformation of $A_0.$ But it
is a domain, while $A_0$ is not and therefore the $\hskip.1cm \Crm [[t]]$
algebras $\tilde{A}_t$ and $A_0 [[t]]$ cannot be isomorphic:
the deformation is {\it nontrivial}.

Similarly $A_t$ and $A_0$ cannot be isomorphic for $t \notin 2 \pi {\bf Q}.$
Furthermore, $\tilde{A}_{t_0 + t}$ is a non trivial deformation of $A_{t_0}$
because the Casimir element $Q_t = F'F+SC+S^2 {\rm cos}t$ takes different
values in $A_{t_0}$ and $A_{t_0+t}$ : in the $(2k+1)$-dimensional
representation its value is sin$(k t )$sin $(k+1) t / {\rm sin}^2 t.$
Therefore, in contradistinction with the other models, the Jimbo models
are not rigid.

{\bf e. Topological quantum double.} Now that we have good models with a nice
duality between them, it is possible to have a good formulation of the quantum
double. To this effect we shall consider ${\cal H}_t (G) \bar{\otimes}
{\cal A}_t (G)$ (with inductive tensor product topology);
 its dual is ${\cal A}_t (G) \hat{\otimes} {\cal H}_t (G)$
 (with the projective tensor product topology). Similarly we can consider
 $C^\infty_t (G) \bar{\otimes} {\cal D}'_t (G).$ The following is true [B2].

{\bf Theorem 2.} \it Let $A$ denote ${\cal A} (G)$ or ${\cal D}' (G)$ or their
deformed versions, and $H$ denote ${\cal H} (G)$ or $C^\infty (G)$ or their
deformed versions. Then the double is $D(A) = A^* \bar{\otimes} A = H
\bar{\otimes} A,$  and its dual is $D(A)^* = A \hat{\otimes} A^* = A
\hat{\otimes} H.$ We have $D(A)^{**} = D(A),$ and these algebras are rigid. \rm
\gsaut
\centerline  {\bf REFERENCES }
\sevenrm
\psaut
\noindent [B{\sevenit ea}] F. Bayen, M. Flato, C. Fronsdal, A. Lichnerowicz
and D. Sternheimer. Lett. Math. Phys. {\sevenbf 1} (1977), p. 521-530.
Ann. Phys. {\sevenbf 111} (1978), 61-151.
\psaut
\noindent [B] F. Berezin. Math. USSR Izv. {\sevenbf 8} (1974), 1109-1165.
\psaut
\noindent [B1] P. Bonneau. Lett. Math. Phys. {\sevenbf 26} (1992), 277-280.
\psaut
\noindent [B2] P. Bonneau. Reviews in Math. Phys. {\sevenbf 6} (1994),
305-318.
\psaut
\noindent [BFGP] P. Bonneau, M. Flato, M. Gerstenhaber and G. Pinczon.
Commun. Math. Phys. {\sevenbf 161} (1994), 125-156.
\psaut
\noindent [C] A. Connes. {\sevenit G\'eom\'etrie non commutative}.
Inter\'editions, Paris (1990).
(English expanded version, preprint IHES/M/93/12, March 1993,
to be published by Academic Press).
\psaut
\noindent [CFS] A. Connes, M. Flato and D. Sternheimer. Lett. Math. Phys.
{\sevenbf 24} (1992), 1-12.
\psaut
\noindent [CM] A. Connes and H. Moscovici. Topology {\sevenbf 20} (1990),
345-388.
\psaut
\noindent [Di] J. Dito. Lett. Math. Phys. {\sevenbf 27} (1993), 73-80.
\psaut
\noindent [D1] V. Drinfeld. Proc. ICM86, vol. 1. Amer Math. Soc. (1987),
798-820.
\psaut
\noindent [D2] V. Drinfeld. Leningrad Math. J. {\sevenbf 1} (1990),
1419-1457.
\psaut
\noindent [EW] C. Emmrich and A. Weinstein. Berkeley Preprint (Nov. 1993).
\psaut
\noindent [FRT] L.D. Faddeev, N.Y. Reshetikhin and L.A. Takhtajan.
Leningrad Math. J., {\sevenbf 1} (1990), 193-226.
\psaut
\noindent [F] B. Fedosov. In "Some Topics of Modern Mathematics and
their applications to problems of Math. Phys." (in Russian) (1985), 129-136.
Sov. Phys. Dokl. {\sevenbf 34} (1989), p. 318-321; J. Diff. Geom. (in press).
\psaut
\noindent [FLS] M. Flato, A. Lichnerowicz and D. Sternheimer. Compositio
Mathematica {\sevenbf 31} (1975), 47-82; C.R. Acad. Sci. Paris {\sevenbf 279}
(1974), 877-881.
\psaut
\noindent [FLuS] M. Flato, Z.C. Lu and D. Sternheimer.
Foundations of Physics {\sevenbf 23} (1993), 587-598.
\psaut
\noindent [FS1] M. Flato and D. Sternheimer. Li\`ege (1977) lectures,
in: {\sevenit Harmonic Analysis and Representations of Semisimple
Lie Groups}, J.A. Wolf, M. Cahen and M. De Wilde (eds.); MPAM vol. 5,
pp. 385-448. D. Reidel, Dordrecht (1980).
\psaut
\noindent [FS2] M. Flato and D. Sternheimer. Lett. Math. Phys.
{\sevenbf 22} (1991), p. 155-160.
\psaut
\noindent [G] M. Gerstenhaber. Ann. Math. {\sevenbf 79} (1964), 59-103.
\psaut
\noindent [GS] M. Gerstenhaber and S.D. Schack. Contemp. Math.
{\sevenbf 134} (1992), 51-92.
\psaut
\noindent [Gr] A. Groenewold. Physica {\sevenbf 12} (1946), 405-460.
\psaut
\noindent [Gu] S. Gutt. Lectures at ICTP Workshop (Trieste, March 1993).
\psaut
\noindent [J] M. Jimbo. Lett. Math. Phys. {\sevenbf 10} (1985), 63-69.
\psaut
\noindent [KR] P.P. Kulish and N.Y. Reshetikhin. J. Sov. Math.
{\sevenbf 23} (1983), 24-35. (In Russian: Zap. Nauch. Sem. LOMI
 {\sevenbf 101} (1981), 101-110).
\psaut
\noindent [LDW] P. Lecomte and M. De Wilde. Lett. Math. Phys.
{\sevenbf 7} (1983), 487-496; Note di Mat. X (in press).
\psaut
\noindent [L] A. Lichnerowicz. Ann. Inst. Fourier {\sevenbf 32}
(1982), 157-209.
\psaut
\noindent [Lu] Z.C. Lu. J. Math. Phys. {\sevenbf 33} (1992), 446-453.
\psaut
\noindent [M] J.E. Moyal. Proc. Cambridge Phil. Soc. {\sevenbf 45} (1949),
99-124.
\psaut
\noindent [NT] R. Nest and B. Tsygan. {\sevenit Algebraic Index Theorem}
to appear in Commun. Math. Phys.
\psaut
\noindent [NV] O.M. Neroslavsky and A.T. Vlasov. C.R. Acad. Sc. Paris
{\sevenbf 292} I  (1981), 71-76.
\psaut
\noindent [OMY] H. Omori, Y. Maeda and A. Yoshioka. Adv. in Math.
{\sevenbf 85} (1991), 225-255; Lett. Math. Phys. {\sevenbf 26} (1992),
 285-294.
\psaut
\noindent [T] L. Takhtajan. Springer Lecture Notes in Physics
{\sevenbf 370} (1990), 3-28.
\psaut
\noindent [Tr] F. Tr\`eves.{\sevenit Topological vector spaces,
distributions and kernels.} Acad. Press. (1967).
\psaut
\noindent [V] J. Vey. Comment. Math. Helv. {\sevenbf 50} (1975), 421-454.
\psaut
\noindent [W] H. Weyl. Z. Physik {\sevenbf 46} (1927), 1-46.
\psaut
\noindent [Wi] E.P. Wigner. Phys. Rev. {\sevenbf 40} (1932), 749-759.
\end